\documentstyle[prl,aps,twoside]{revtex}
\oddsidemargin=\evensidemargin
\begin{document}
\twocolumn[\hsize\textwidth\columnwidth\hsize\csname@twocolumnfalse%
\endcsname
\title{Pseudo-Goldstone boson effects in $t\bar{t}$ productions at
high energy hadron colliders and testing technicolor models}
\bigskip\bigskip\bigskip
\author{Ling Zhang$^b$~~~~~~Xue-Lei Wang$^{abc}$~~~~~~Yu-Ping
Kuang$^{ab}$~~~~~~Hong-Yi Zhou$^{ab}$}
\address{a. China Center of Advanced Science and Technology (World
Laboratory), P.O.Box 8730, Beijing 100080, China\\
b. Department of Physics, Tsinghua University, Beijing 100084,
China\footnote{Mailing address.}\\
c. Physics Department, Henan Normal University, Xinxiang 453002, Henan,
China}
\bigskip\bigskip\bigskip
\date{TU-HEP-TH-99/103}
\maketitle
\null\vspace{0.5cm}
\begin{abstract}
We study the top quark pair production process $pp(\bar{p})
\to t\bar{t}$ in various kinds of technicolor (TC) models at the Fermilab
Tevatron Run II and the CERN LHC. The s-channel neutral pseudo-Goldstone 
bosons (PGBs) contribute dominantly to the production amplitudes from its 
coupling to the gluons through the triangle loops of techniquarks and the top 
quark. Cross sections in different TC models with s-channel PGB contributions 
are calculated. It is shown that the PGB effects can be experimentally tested 
and different TC models under consideration can be distinguished at the LHC. 
Therefore, the $pp\to t\bar{t}$ process at the LHC provides feasible tests of 
technicolor models.
\end{abstract}
\vspace{0.2cm}\hspace{1.84cm}
PACS number(s): 12.15.Lk, 12.60.Nz, 13.30.Eg
\bigskip\bigskip\bigskip\bigskip\bigskip
]
\begin{center}
{\bf I. INTRODUCTION}
\end{center}

Understanding the mechanism of electroweak symmetry breaking (EWSB) is
one of the most important problems in current particle physics and will
be studied experimentally at present and future high energy colliders, for 
instance the CERN LEP2 and LHC, the Fermilab Tevatron Run II, and the future 
$e^+e^-$ linear colliders. The standard model (SM) Higgs boson has not been 
found yet, and the Higgs sector in the SM suffers from the well-known problems 
of {\it triviality} and {\it unnaturalness}. So that the EWSB mechanism is 
possibly related to new physics beyond the SM. Two main candidates of
new physics related to the EWSB mechanism are supersymmetry and
dynamical EWSB mechanism, for example various technicolor (TC) models. 

TC models are based on new strong interaction dynamics which is
difficult to deal with. However, TC models contain certain new particles, such
as new heavy gauge bosons and new resonances.
It is feasible to test TC models via processes including the
contributions of these new particles. On the other hand, The top quark 
is the heaviest particle yet discovered. The directly measured top quark mass
is $m_t=174.3\pm 3.2\pm 4.0$ GeV \cite{Heinson} which is close to the EWSB 
scale $v=246$ GeV, so that the effective Yukawa coupling of the top quark is
of the order of 1. Thus TC models may be detected through studying new particle
contributions to top quark production processes at future high energy 
colliders. Study of the contributions of $s$-channel new heavy gauge bosons and
new vector-resonances (color-singlet, color-octet, or hybrid) in TC models of 
masses ranging from 600 GeV to 1 TeV to $t\bar t$ productions at the Tevatron
has been given in Ref.\cite{HP}, which shows that the effects are 
experimentally detectable.
Another characteristic feature of the TC models is that most TC models predict 
certain pseudo-Goldstone bosons (PGBs) of masses below 1 TeV,
and the properties of the PGBs are different in different models. Therefore,
studying the effects of the PGB contributions in $t\bar t$ productions
at high energy colliders can serve as good tests of TC models. The effect 
of color-octet technipions $\Pi^{0a}~(a=1,\cdots,8)$ on $t\bar t$ production 
at the Fermilab Tevatron has been studied in Ref.\cite{EL} and it shows that 
$\Pi^{0a}$ can make important contributions via the gluon fusion process 
$gg\to\Pi^{0a}\to t\bar t$ due to the large PGB-gluon-gluon coupling 
contributed by the techniquark triangle-loop, and such effect can be tested 
by measuring the differential cross section. A more complete study of the PGB 
effects in the $t\bar t$ productions at the Tevatron and the LHC in the 
topcolor-assisted multiscale TC model has been studied in Ref.\cite{hadtt}
in which the contributions from the color-singlet technipion and the
top-pion are included as well, and the total effects are shown to be large 
enough to be experimentally detected. 
In Ref.\cite{gamgamtt}, the $\gamma\gamma\to t\bar{t}$ process at the
future photon collider in various TC models (with and without topcolor) is 
studied. It is shown that the $s$-channel PGB contributions are dominant, and 
the results show that different TC models can be experimentally distinguished 
at the future photon colliders. In this paper, we extend the study of 
Ref.\cite{hadtt} to include various kinds of typical TC models as studied in 
Ref.\cite{gamgamtt} and examine whether those typical TC models can also be 
experimentally distinguished at the future hadron colliders. Since 
there can be $s$-channel color-octet PGBs contributing to the $t\bar{t}$ 
productions at the hadron colliders, the present case is different from 
that in Ref.\cite{gamgamtt}. Our calculation will show that,
with the expected systematic errors in the $t\bar{t}$ cross section
measurements at the LHC, experimentally testing and distinguishing different 
kinds of TC models are possible by measuring the $t\bar{t}$ production cross 
section and the invariant mass distribution at the LHC. Therefore, the 
$t\bar t$ production process provides feasible tests of technicolor models.

This paper is organized as follows. In Sec. II, we present the
calculation of the production amplitudes for three typical TC models
with and without topcolor.
The numerical results of the production cross sections are presented in
Sec. III, and a concluding remark is given in Sec. IV.

\null\vspace{0.4cm}
\begin{center}
{\bf II. CALCULATION OF THE PRODUCTION AMPLITUDES FOR THREE TYPICAL TC MODELS}
\end{center}

We shall take into account the tree-level SM amplitude ${\cal M}^{SM}_{tree}$
and the PGB contributed amplitude $M_{\Pi}$ ($\Pi$ stands for the
related PGB in this paper) described in Fig. 1. At the
hadron colliders, ${\cal M}^{SM}_{tree}$ mainly contains two parts, namely the
quark fusion amplitude ${\cal M}^{SM}_{tree}(q\bar q\to t\bar t)$ and the gluon
fusion amplitude ${\cal M}^{SM}_{tree}(gg\to t\bar t)$, i.e.
\begin{eqnarray}                                  
{\cal M}^{SM}_{tree}={\cal M}^{SM}_{tree}(q\bar q\to t\bar t)
+{\cal M}^{SM}_{tree}(gg\to t\bar t)\,.
\label{Mtree}
\end{eqnarray}
 For the LHC,
${\cal M}^{SM}_{tree}(gg\to t\bar t)$ dominates, and we shall neglect
${\cal M}^{SM}_{tree}(q\bar q\to t\bar t)$ in the calculation. For the 
Tevatron, although ${\cal M}^{SM}_{tree}(q\bar q\to t\bar t)$ is dominant, the 
interference term between ${\cal M}^{SM}_{tree}(gg\to t\bar t)$ and 
${\cal M}_{\Pi}$ is actually not negligibly small, so that we shall take 
account of both ${\cal M}^{SM}_{tree}(q\bar q\to t\bar t)$ and 
${\cal M}^{SM}_{tree}(gg\to t\bar t)$
in our calculation. For ${\cal M}_{\Pi}$, as in Ref.\cite{gamgamtt}, we take 
the Appelquist-Terning one family walking TC model (Model A) \cite{AT}, the 
original topcolor-assisted technicolor model (Model B) \cite{TC2-1} and the 
topcolor-assisted multiscale technicolor model (Model C) \cite{TC2-2} as three 
typical examples of TC models with and without topcolor to illustrate the 
results. The details will be presented as follows.

\noindent
\subsection{The Appelquist-Terning One Family Walking TC Model (Model A) }

We take this model as a typical example of the improved TC models
without assisted by topcolor. To reduce the value of the oblique correction 
parameter $S$, this model is designed such that the techniquark ($Q$) sector 
respects the custodial $SU(2)$ symmetry, while the technilepton ($L$) sector 
is custodial $SU(2)$ violating, and the vacuum expectation values (VEV's)
of $\bar{Q}Q$ and $\bar{L}L$ are further designed to be $~F_Q\gg F_L$ \cite{AT}. 
The color-singlet would-be Goldstone bosons eaten by $W$ and $Z$
are mainly composed of techiquarks. There are 36 PGBs in this model \cite{AT},
in which the color-singlet PGBs are mainly composed of technileptons which 
are irrelevant to the $s$-channel $t\bar{t}$ production. At the hadron 
colliders, the color-octet PGBs $\Pi ^{0a}~(a=1,...,8)$ composed of 
techniquarks can contribute to the $s$-channel $t\bar{t}$ productions via the
techniquark and top quark triangle loops [cf. Fig. 1]. This is the main 
difference between the present case and the $\gamma\gamma\to t\bar{t}$ case in 
Ref.\cite{gamgamtt}. The decay constant of the color-octet PGBs is
$F_\Pi=123$ GeV \cite{AT}.  The masses of $\Pi^{0a}$ are model-dependent.
Following Ref.\cite{AT}, we take $M_{\Pi^{0a}}$ in the range 
$400~{\rm GeV}\alt M_{\Pi^{0a}} \alt 500$ GeV.

Since the techniquark $Q$ is very heavy, the triangle loop in Fig. 1(a) can be
simply evaluated by the Adler-Bell-Jackiw anomaly \cite{anomaly}, and the
general form of which is \cite{Lubicz,DE}
\begin{eqnarray}                               
\frac{S_{\Pi^aB_1B_2}}{4\pi^2 F_{\pi}} \epsilon_{\mu\nu\lambda\rho}
k_1^{\lambda}k_2^{\rho}\,,
\label{ABJ}
\end{eqnarray}
where $B_1$ and $B_2$ denote the two gauge fields which, in our case, are the
two gluons $g_b$ and $g_c$ with the color indices $b$ and $c$, respectively.
The factor $S_{\Pi^a g_bg_c}$ can be easily obtained from the formulae
in Ref.\cite{Lubicz,DE,hadtt}, and it is\footnote{Here, and in (\ref{t-loop}),
(\ref{J-Pia}) and (\ref{S(B)}), we have corrected some typos in 
Ref.\cite{hadtt}.}
\begin{eqnarray}                               
S_{\Pi^{0a}g_bg_c} =\frac{1}{\sqrt{2}}g_s^2 N_{TC}d_{abc}\,,
\label{S_Piagg}
\end{eqnarray}
where $d_{abc}$ is the symmetric tensor in the color $SU(3)_c$ group.

The evaluation of the triangle loop in Fig. 1(b) needs more consideration. The 
top quark is not heavy enough for the validity of simply using the 
Adler-Bell-Jackiw anomaly. Correction of the $m_t$ effect has to be
taken into account. This has been calculated in Ref.\cite{J}, and the
result is
\begin{eqnarray}
-i\frac{C_t g_s^2}{8\pi^2 F_{\pi}}\frac{d_{abc}}{2} J(R_{\hat{s}})
\epsilon_{\mu \nu \lambda \rho} k_1^{\lambda} k_2^{\rho}\,,              
\label{t-loop}
\end{eqnarray}
where $C_t$ is a model-dependent coupling constant which is expected to be
of order 1 \cite{TC2-2,Lubicz,RS}, $\hat{s}$ is the center-of mass energy of 
the $t\bar t$ system and $J(R_{\hat{s}})$ is defined as \cite{J}
\begin{eqnarray}
J(R_{\hat{s}})&\equiv& -\frac{1}{R^2_{\hat{s}}}
\int_0^1 \frac{dx}{x(1-x)}\nonumber\\
&&\times \ln[1 - R^2_{\hat{s}} x(1-x) ]\,,           
\label{J-Pia}
\end{eqnarray}
with $R_{\hat{s}} \equiv \sqrt{\hat s}/m_t$.

Combining (\ref{ABJ}), (\ref{S_Piagg}) and (\ref{t-loop}), we obtain the 
production amplitude for Fig. 1
\begin{eqnarray}                                                      
\displaystyle
 {\cal M}^{(A)}_{\Pi^{0a}}&=& \frac{C_tm_tg_s^2[N_{TC}+\frac{1}{2\sqrt{2}}
 C_tJ(R_{\hat s})]d_{abc}}{4\sqrt{2}\pi^2F^2_{\Pi}[\hat{s}-M^2_{\Pi^{0a}}
+iM_{\Pi{0a}}\Gamma_{\Pi^{0a}}]}
\nonumber\\
&&\times(\bar{t}\gamma_5\frac{\lambda_a}{2}t)\epsilon_{\mu\nu\lambda\rho}
\varepsilon_1^{\mu}\varepsilon_2^{\nu}k_1^{\lambda}k_2^{\rho}\,,
\end{eqnarray}
where $\Gamma_{\Pi^{0a}}$ is the total width of $\Pi^{0a}$ which
has been given in Ref.\cite{hadtt}. The total production amplitude is then
\begin{eqnarray}                                                     
{\cal M}^{(A)}={\cal M}^{SM}_{tree}+ {\cal M}^{(A)}_{\Pi^{0a}}\,.
\label{ATamplitude}
\end{eqnarray}

\subsection{The Original Topcolor-Assisted Technicolor Model (Model B)}

Since the original topcolor-assisted technicolor model (Model B) was
proposed \cite{TC2-1}, there have been refinements of the model to make
it more realistic \cite{TC2-1'}. In the present study, we are only
interested in the characteristic PGB effects of this kind of model in 
$t\bar t$ productions which do not concern the subtleties of the
refinements, so that we simply take the original Model B as a typical
example of this kind of model in our calculation.
In this model, the TC sector is taken to be the standard extended
technicolor model in which there are 60 TC PGBs with the decay
constant $F_\Pi\approx 120$ GeV\footnote{This is slightly smaller than
the usual value $F_\Pi=123$ GeV in the extended technicolor model since, in 
the topcolor-assisted technicolor model, the total vacuum expectation value is 
also contributed by the topcolor sector.}, and both the 
the color-octet PGB $\Pi^{0a}$ and the color-singlet PGB $\Pi^0$ contribute to 
the $t\bar t$ production. As in Model A, we take 
$400\alt M_{\Pi^{0a}}\alt 500$ GeV. The mass of $\Pi^0$
is lighter, say around $150$ GeV \cite{ETC}. The coupling of $\Pi^0$ to
gluons via the techniquark and top quark triangle loops is
decscribed by \cite{Lubicz,hadtt}
\begin{eqnarray}                                          
\displaystyle
S^{(B)}_{\Pi^0 g_b g_c}&=&\frac{1}{2\sqrt{3}}g^2_s\delta_{bc}N_{TC}\nonumber\\
&&+\frac{1}{\sqrt{2}}g^2_sJ(R_{\hat s})\delta_{bc}\,,
\label{S(B)}
\end{eqnarray}
where the first term is from the techniquark loop and the second term
is from the top quark loop.

There is a topcolor sector in this model responsible for causing the main part 
of the top quark mass. In the topcolor sector, there is a PGB called top-pion 
$\Pi_t$ with decay constant $F_{\Pi_t}=50$ GeV. The mass $M_{\Pi_t}$
was first estimated as around $200$ GeV in the original paper \cite{TC2-1}. 
However, recent phenomenological analyses up to one-loop calculations show 
that the LEP/SLD precision data of $R_b$ give severe constraint on the value 
of $M_{\Pi_t}$ due to the large negative contribution to $R_b$ from the 
corrections related to $\Pi_t$, and it requires $M_{\Pi_t}$ to be of the order 
of 1 TeV \cite{BK,LT}. However, as is pointed out in Ref.\cite{BK}, such a 
constraint can only be regarded as a rough estimate since higher order
corrections related to $\Pi_t$ may be substantial due to the large 
$\Pi_t-t-\bar b$ coupling. Furthermore, the extended technicolor gauge boson 
contribution to $R_b$, which has been shown to be positive \cite{R_b}, is not 
taken into account in the analyses in Refs.\cite{BK,LT}, and the actual 
constraint on $M_{\Pi_t}$ may be relaxed when such a positive contribution is 
taken into account.
Therefore, to see the $M_{\Pi_t}$-dependence of the cross 
section, we take $M_{\Pi_t}$ to vary in the range $~500~{\rm GeV}\alt
M_{\Pi_t}\alt 1~{\rm TeV}~$ in our calculation. 

The top quark mass $m_t$ comes from two sources in this model. The TC
sector gives rise to a small portion of it, and we call this portion 
$m^\prime_t$. The value of $m^\prime_t$ is model-dependent.
Low energy data, especially the $b\to s\gamma$ experiment, give constraints on 
$m^\prime_t$, and the reasonable range of $m^\prime_t$ is about 
$5~{\rm GeV}\alt m^\prime_t\alt 20~{\rm GeV}$ \cite{TC2-1,Balaji}. The rest 
part of $m_t$, say $m_t-m^\prime_t$ comes from the topcolor sector. Thus the 
couplings of the technipions to the top quark can be written as 
\cite{EL}\cite{Lubicz}
\begin{eqnarray}
\displaystyle
\frac{C_t m_t^\prime}{\sqrt{2} F_\Pi} \Pi^0 (\bar{q} \gamma^5 q)    
\end{eqnarray}
\begin{eqnarray}                                                
\frac{C_t m_t^{\prime}}{F_\Pi} \Pi^{0a} (\bar{q} \gamma^5 \frac{\lambda^a}{2} 
q)
\end{eqnarray}
where $\lambda^a$ is the Gell-Mann matrix of the color group. 
The interactions of the top-pions with the top quark is \cite{TC2-1,TC2-1'}
\begin{eqnarray}
\displaystyle
\frac{m_t - m_t^{\prime}}{\sqrt{2} F_{\Pi_t}} [\bar{t} \gamma_5 t \Pi_t^0 +
\frac{i}{\sqrt{2}} \bar{t} (1 - \gamma_5) b \Pi_t^+ \nonumber\\
+ \frac{i}{\sqrt{2}}\bar{b} (1 + \gamma_5) t \Pi_t^-]\,.               
\end{eqnarray}

With these couplings, the PGB contributed production amplitudes in this
model described in Fig. 1 are
\begin{eqnarray}                                                      
\displaystyle
 {\cal M}^{(B)}_{\Pi^{0a}}&=& \frac{C_tm_t^\prime g_s^2[N_{TC}+\frac{1}
{2\sqrt{2}}C_t J(R_{\hat s})]
d_{abc}}{4\sqrt{2}\pi^2F^2_{\Pi}[\hat{s}-M^2_{\Pi^{0a}}
+iM_{\Pi{0a}}\Gamma_{\Pi^{0a}}]}
\nonumber\\
&&\times(\bar{t}\gamma_5\frac{\lambda_a}{2}t)\epsilon_{\mu\nu\lambda\rho}
\varepsilon_1^{\mu}\varepsilon_2^{\nu}k_1^{\lambda}k_2^{\rho}\,,\label{MPi0a}
\end{eqnarray}
\begin{eqnarray}                                                 
\displaystyle
{\cal M}^{(B)}_{\Pi^0}&=&\frac{C_tm_t^{\prime}g_s^2[N_{TC}+\frac{\sqrt{6}}{2}
C_t J(R_{\hat s})]\delta_{bc}}
{8\sqrt{6}\pi^2F^2_{\Pi}[\hat{s}-M^2_{\Pi^0}+iM_{\Pi^0}\Gamma_{\Pi^0}]}
\nonumber\\
&&\times (\bar{t}\gamma_5t)\epsilon_{\mu\nu\lambda\rho}\varepsilon_1^
{\mu}\varepsilon_2^{\nu}k_1^{\lambda}k_2^{\rho}\,,\label{MPi0}
\end{eqnarray}
\begin{eqnarray}                                                
\displaystyle
{\cal M}^{(B)}_{\Pi_t}&=&\frac{(m_t-m_t^{\prime})g_s^2J(R_{\hat s})\delta_{bc}}
{8\sqrt{6}\pi^2F^2_{\Pi_t}[\hat{s}-M^2_{\Pi_t^0}+iM_{\Pi_t^0}\Gamma_{\Pi_t^0}]}
\nonumber\\
&&\times (\bar{t}\gamma_5t)\epsilon_{\mu\nu\lambda\rho}\varepsilon_1^{\mu}
\varepsilon_2^{\nu}k_1^{\lambda}k_2^{\rho}\,,\label{M^B_Pi_t}
\end{eqnarray}
where $\Gamma_{\Pi^0}$ and $\Gamma_{\Pi_t}$ are, respectively, the total 
widths of $\Pi^0$ and $\Pi_t$ given in Ref.\cite{hadtt}.
The total production amplitude in this model is then
\begin{eqnarray}                                             
{\cal M}^{(B)}={\cal M}^{SM}_{tree}+{\cal M}^{(B)}_{\Pi^{0a}}
+{\cal M}^{(B)}_{\Pi^0}+{\cal M}^{(B)}_{\Pi_t}\,.
\end{eqnarray}
Compared with ${\cal M}^{(A)}$, the amplitude ${\cal M}^{(B)}$ contains two 
extra terms ${\cal M}^{(B)}_{\Pi^0}$ and ${\cal M}^{(B)}_{\Pi_t}$. As we shall 
see later that this makes Model A and Model B experimentally distinguishable at
the LHC.

\subsection{The Topcolor-Assisted Multiscale Technicolor Model (Model C)}

The topcolor-assisted multiscale technicolor model (Model C) 
\cite{TC2-2,EL,hadtt} is different from Model B by its extended technicolor 
sector which is taken to be the multiscale technicolor model \cite{MTC}.
In this model, the value of the decay constant $F_\Pi$ is $F_\Pi=40$ GeV 
rather than $120$ GeV, and the technipion $\Pi^0$ is almost 
composed of pure techniquarks (ideal mixing) \cite{TC2-2} which leads to
\begin{eqnarray}                                            
\displaystyle
S^{(C)}_{\Pi^0 g_b g_c}=\frac{1}{\sqrt{3}}N_{TC}\delta_{bc}\,.
\label{S(C)}
\end{eqnarray}
Then the production amplitudes in Model C is
\begin{eqnarray}                                            
{\cal M}^{(C)}={\cal M}^{SM}_{tree}+{\cal M}^{(C)}_{\Pi^{0a}}
+{\cal M}^{(C)}_{\Pi^0}+{\cal M}^{(C)}_{\Pi_t}\,,
\end{eqnarray}
with
\begin{itemize}
\item ${\cal M}^{(C)}_{\Pi_t}={\cal M}^{(B)}_{\Pi_t}$;
\item the formula for ${\cal M}^{(C)}_{\Pi^0}$ differs from that for
${\cal M}^{(B)}_{\Pi^0}$ by a factor of $2$;
\item the value of $F_\Pi$ in ${\cal M}^{(C)}_{\Pi^{0a}}$ and 
${\cal M}^{(C)}_{\Pi^0}$ is $F_\Pi=40$ GeV rather than $123$ GeV.
\end{itemize}
The smallness of the value of $F_\Pi$ and the ideal mixing of $\Pi^0$
in model C enhance the technicolor PGB contributions in $t\bar t$ production
relative to the top-pion contribution. As we shall see later that this
makes Model C experimentally distinguishable from Model B and Model A.

\null\vspace{0.4cm}
\begin{center}
{\bf III. CROSS SECTIONS AND NUMERICAL RESULTS}
\end{center}

We take the method in Ref.\cite{helicity} to do the numerical calculation. 
Once the elementary cross section $\hat \sigma$ is calculated at the 
parton-level, the total cross section $\sigma$ can be obtained by folding 
$\hat \sigma$ with the parton distribution functions $f^{p(\bar p)}_i(x_i,Q)$ 
\cite{EHLQ}
\begin{eqnarray}                                               
\sigma(pp(\bar{p})\to t\bar{t})&=&\sum\limits_{ij}\int dx_idx_jf_i^{(p)}(x_i,Q)
f_j^{(p(\bar{p}))}(x_j,Q)\nonumber\\
&&\times\hat{\sigma}(ij\to t\bar{t})
\end{eqnarray}
where $i$ and $j$ stand for the partons $g$, $q$ and $\bar{q}$; $x_i$ is the 
fraction of longitudinal momentum of the proton (antiproton) carried by the 
$i$th parton; $Q^2\approx \hat s$; and $f_i^{(p(\bar{p}))}$ is the parton 
distribution function in the proton (antiproton). In this paper, we take the 
MRS setA$^\prime$ parton distribution for $f_i^{p(\bar{p})}$ \cite{MRS}. To take 
account of the QCD corrections, we shall multiply the obtained cross section 
by a factor of 1.6 \cite{LSN} as what was done in Ref.\cite{hadtt}. The
values of the tree-level SM cross section $\sigma_0$ at the $\sqrt{s}=2$ TeV
Tevatron and the $\sqrt{s}=14$ TeV LHC are, respectively
\begin{eqnarray}                                             
&&{\rm Tevatron}:~~~~~~~~~~~~~~\sigma_0=8.02~{\rm pb}\,,\nonumber\\
&&{\rm LHC}:~~~~~~~~~~~~~~~~~~~~\sigma_0=826~{\rm pb}\,.
\label{sigma_0}
\end{eqnarray}
In the numerical calculations, we take $\alpha_s(\sqrt{\hat s})$ the same as
that in the MRS set A$^\prime$ parton distributions, $m_t=174$ GeV,
and we simply take the technicolor model parameter $C_t=1$. In the
following analysis, we consider the one-year-run integrated luminosities
for the Tevatron Run II and the LHC
\begin{eqnarray}                                       
{\rm Tevatron}:~~~~~~~~\int {\cal L}dt&=&2~{\rm fb}^{-1}\,,\nonumber\\
{\rm LHC}:~~~~~~~~~~~~~\int {\cal L}dt&=&100~{\rm fb}^{-1}\,,
\label{luminosity}
\end{eqnarray}
and assume a $10\%$ detecting efficiency. 

The obtained total production
cross sections can be compared with the recently measured $t\bar t$ 
production cross sections by the CDF Collaboration and the D0 Collaboration 
\cite{Heinson}
\begin{eqnarray}                                        
{\rm CDF}:~~~~~~~~~~\sigma(p\bar p\to t\bar t)&=&10.1\pm 1.9^{+4.1}_{-3.1}
~{\rm pb}\,,\nonumber\\
{\rm D0}:~~~~~~~~~~~\sigma(p\bar p\to t\bar t)&=&7.1\pm 2.8\pm 1.5~{\rm pb}\,.
\label{sigmatt}
\end{eqnarray}
The data in (\ref{sigmatt}) can serve as a constraint on the parameters in the 
TC models.

\subsection*{A. Results of Model A}

In Table I, we list the results of the cross sections at the Tevatron Run II 
and the LHC in Model A with $M_{\Pi^{0a}}$ varying from 400 GeV to 500 GeV. We 
see from Table I that the values of $\sigma^{(A)}_{t\bar t}$ for the Tevatron 
are consistent with the recent CDF and D0 measurements (\ref{sigmatt}). The 
relative technicolor corrections to the SM tree-level cross section $\sigma_0$
are $\Delta\sigma^{(A)}/\sigma_0\approx (10-36)\%$ for the Tevatron

\null\vspace{0.4cm}
\noindent
{\small Table I.  Cross sections in Model A at the $\sqrt{s}=2$ TeV Tevatron
and the $\sqrt{s}=14$ TeV LHC with $M_{\Pi^{0a}}$ varying from 400 GeV to 500 
GeV. $\sigma_0$ denotes the SM tree-level cross section,
$\Delta\sigma^{(A)}$ denotes the correction to $\sigma_0$,
and $\sigma^{(A)}_{t\bar{t}}=\sigma_0+\Delta\sigma^{(A)}$ is the total cross 
section. All masses are in GeV.}
\begin{center}
\doublerulesep 0.5pt
\tabcolsep 0.5pt
\begin{tabular}{c cc cc}
\hline\hline
& $~~~~~~~~~~$Tevatron& &$~~~~~~~$LHC& \\
\hline
$M_{\Pi^{0a}}$&$\Delta\sigma^{(A)}(pb)$&$\sigma^{(A)}_{t\bar{t}}(pb)$ &
$~~~~\Delta\sigma^{(A)}(nb)$ & $~~\sigma^{(A)}_{t\bar{t}}(nb)~~$\\
\hline
400 & 2.92 & 10.94 ~~& 1.36 & 2.19 \\
450 & 1.54 & 9.56 ~~& 1.04 & 1.87 \\
500 & 0.84 & 8.86 ~~& 0.81 & 1.63 \\
\hline\hline
\end{tabular}
\end{center}
\vspace{0.4cm}

\noindent
and $\Delta\sigma^{(A)}/\sigma_0\approx (98-165)\%$ for the LHC, which are 
quite large due to the $\Pi^{0a}$ resonance effects. The relative corrections 
are much larger than those in the $\gamma\gamma\to t\bar t$ process given in 
Ref.\cite{gamgamtt} because of the existence of the $\Pi^{0a}$ contribution
at the hadron colliders.
With the integrated luminosities in (\ref{luminosity}) and assuming a $10\%$ 
detecting efficiency, we see from Table I that Model A predicts around
2000 $t\bar t$ events at the Tevatron and around $2\times 10^7~ t\bar t$ events
at the LHC. The statistical uncertainty at the $95\%$ C.L. in the case of the 
Tevatron is then around $4\%$ which is about the same level as the expected 
systematic error of the $t\bar t$ cross section measurement ($\sim 5\%$ 
\cite{Liss}), and the statistical uncertainty in the case of the LHC is 
around $4\times 10^{-4}$ which is much smaller than the expected systematic 
error ($\sim {\rm few}\%$ \cite{Liss}). The relative corrections 
$\Delta\sigma^{(A)}/\sigma_0$ from Table I are all larger than the above
uncertainties and thus {\it these events are all experimentally
detectable at both the Tevatron and the LHC}. To illustrate the resonances, 
we further plot the $t\bar t$ invariant mass distributions for
$M_{\Pi{0a}}=400$ GeV at the Tevatron and the LHC in Fig. 2(a) and Fig. 2(b), 
respectively. The resonance effects at $M_{\Pi^{0a}}$ can be clearly
seen. Comparing Fig. 2(a) with the new vector resonances (with
the width about $20\%$ of the mass) shown in Ref.\cite{HP}, we see that the 
$\Pi^{0a}$ resonance is sharper.

\subsection*{B. Results of Model B}

The results of the cross sections in Model B at the Tevatron are listed in 
Table II. Since $M_{\Pi^0}$ is much lower than the $t\bar t$ threshold, there 
is almost no $\Pi^0$ resonance effect, so that we simply take a typical value 
$M_{\Pi^0}=150$ GeV in the calculation. To see the resonance effects of 
$\Pi^{0a}$ and $\Pi_t$ with various values of $M_{\Pi^{0a}}$ and $M_{\Pi_t}$, 
we take their masses varying in the ranges $~400~{\rm GeV}\alt M_{\Pi^{0a}}
\alt 500~{\rm GeV}~$ and $~500~{\rm GeV}\alt M_{\Pi_t}\alt 1~{\rm TeV}$, 
respectively. For the parameter $m^\prime_t$, we take two typical values 
$m^\prime_t=5$ GeV (denoted by the superscript $i=1$) and $m^\prime_t=15$ GeV 
(denoted by the superscript $i=2$), and the cross sections with these
two values of $m^\prime_t$ are denoted by $\sigma^{(B1)}_{t\bar t}$ 

\null\vspace{0.4cm}
\noindent
{\small Table II. Cross sections in Model B at the $\sqrt{s}=2$ TeV Tevatron. 
$\Delta\sigma^{(Bi)}$ denotes the correction to the SM tree-level cross
section $\sigma_0$, and $\sigma^{(Bi)}_{t\bar{t}}=\sigma_0
+\Delta\sigma^{(Bi)}$ is the total cross section. The superscript $i$
denotes the two cases of $m_t^ {\prime}=5$ GeV ($i=1$) and $m_t^\prime=15$ GeV
($i=2$). All masses are in GeV, and all cross sections are in pb.}
\begin{center}
\doublerulesep 0.5pt
\tabcolsep 5pt
\begin{tabular}{cccccc}\hline\hline
$M_{\Pi_t^0}$  & $M_{\Pi^{0a}}$  & $\Delta\sigma^{(B1)}$ &
$\sigma^{(B1)}_{t\bar{t}}$ & $\Delta\sigma^{(B2)}$ & $\sigma^{(B2)}
_{t\bar{t}}$ \\
\hline
500 & 400 & 0.13 & 8.15 & 0.47 & 8.49 \\
500 & 450 & 0.11 & 8.13 & 0.29 & 8.31 \\
500 & 500 & 0.09 & 8.11 & 0.18 & 8.20 \\
600 & 400 & 0.10 & 8.12 & 0.44 & 8.46 \\
600 & 450 & 0.08 & 8.09 & 0.26 & 8.28 \\
600 & 500 & 0.06 & 8.08 & 0.15 & 8.17 \\ 
700 & 400 & 0.08 & 8.10 & 0.42 & 8.44 \\
700 & 450 & 0.06 & 8.08 & 0.24 & 8.26 \\
700 & 500 & 0.05 & 8.06 & 0.14 & 8.15 \\
800 & 400 & 0.07 & 8.09 & 0.42 & 8.43 \\
800 & 450 & 0.05 & 8.07 & 0.24 & 8.25 \\
800 & 500 & 0.04 & 8.06 & 0.13 & 8.15 \\
900 & 400 & 0.07 & 8.08 & 0.41 & 8.43 \\
900 & 450 & 0.05 & 8.06 & 0.23 & 8.25 \\
900 & 500 & 0.03 & 8.05 & 0.13 & 8.14 \\
1000 & 400 & 0.06 & 8.08 & 0.41 & 8.42 \\
1000 & 450 & 0.04 & 8.06 & 0.23 & 8.24 \\
1000 & 500 & 0.03 & 8.05 & 0.12 & 8.14 \\
\hline\hline
\end{tabular}
\end{center}

\vspace{0.4cm}
\noindent
and $\sigma^{(B2)}_{t\bar t}$, respectively. From the values of 
$\sigma^{(B1)}_{t\bar t}$ and $\sigma^{(B2)}_{t\bar t}$ in Table II, we
see that they are consistent with the recent CDF and D0 measurements 
(\ref{sigmatt}). We know that the width of a heavy $\Pi_t$ is 
rather large due to the largeness of $(m_t-m^\prime_t)/F_{\Pi_t}$ (the
smallness of $F_{\Pi_t}$), thus  
the cross sections depend more sensitively on $M_{\Pi^{0a}}$ than on 
$M_{\Pi_t}$ as we see in Table II. Moreover, the $\Pi^{0a}$ couplings are 
proportional to $m^\prime_t$, while the $\Pi_t$ couplings are
proportional to $m_t-m^\prime_t$. The former is much sensitive to
$m^\prime_t$ than the latter does since $m_t\gg m^\prime_t$. Thus the cross 
sections with $m^\prime_t=15$ GeV are all larger than those with 
$m^\prime_t=5$ GeV in Table II. From Table II we see that all relative 
corrections $\Delta\sigma^{(B1)}/\sigma_0$ and $\Delta\sigma^{(B2)}/\sigma_0$ 
are at most $6\%$ which is of the same order as the expected systematic error 
($\sim 5\%$). Hence {\it Model B can hardly be detected at the Tevatron}.

The obtained cross sections in Model B at the LHC are listed in Table III.
Now the relative corrections $|\Delta\sigma^{(B1)}|/\sigma_0$
in Table III are around $(3\--7)\%$ depending on the value of $\Pi_t$.
Since the statistical uncertainty is of the order of $10^{-4}$ as can
be seen from Table III and
\newpage
{\small Table III. Cross sections in Model B at the $\sqrt{s}=14$ TeV LHC. 
$\Delta\sigma^{(Bi)}$ denotes the correction to the SM tree level cross 
sections $\sigma_0$, and $\sigma^{(Bi)}_{t\bar{t}}=\sigma_0
+\Delta\sigma^{(Bi)}$ is the total cross section. The superscript $i$ 
denotes the two cases of $m_t^ {\prime}=5$ GeV ($i=1$) and $m_t^\prime=15$ GeV
($i=2$). All masses are in GeV, and all cross sections are in nb.}
\begin{center}
\doublerulesep 0.5pt
\tabcolsep 5pt
\begin{tabular}{cccccc}
\hline\hline
$M_{\Pi_t^0}$  & $M_{\Pi^{0a}}$  & $\Delta\sigma^{(B1)}$ &
$\sigma^{(B1)}_{t\bar{t}}$ & $\Delta\sigma^{(B2)}$ & $\sigma^{(B2)}
_{t\bar{t}}$ \\
\hline
500 & 400 & 0.06 & 0.89 & 0.23 & 1.05 \\
500 & 450 & 0.06 & 0.88 & 0.19 & 1.02 \\
500 & 500 & 0.05 & 0.88 & 0.15 & 0.98 \\
600 & 400 & 0.05 & 0.87 & 0.21 & 1.04 \\
600 & 450 & 0.05 & 0.87 & 0.18 & 1.01 \\
600 & 500 & 0.04 & 0.87 & 0.14 & 0.97 \\ 
700 & 400 & 0.04 & 0.87 & 0.20 & 1.03 \\
700 & 450 & 0.04 & 0.86 & 0.17 & 1.00 \\
700 & 500 & 0.03 & 0.86 & 0.13 & 0.96 \\
800 & 400 & 0.04 & 0.86 & 0.20 & 1.03 \\
800 & 450 & 0.03 & 0.86 & 0.17 & 0.99 \\
800 & 500 & 0.03 & 0.86 & 0.13 & 0.95 \\
900 & 400 & 0.03 & 0.86 & 0.20 & 1.02 \\
900 & 450 & 0.03 & 0.86 & 0.16 & 0.99 \\
900 & 500 & 0.03 & 0.85 & 0.12 & 0.95 \\
1000 & 400 & 0.03 & 0.86 & 0.19 & 1.02 \\
1000 & 450 & 0.03 & 0.85 & 0.16 & 0.99 \\
1000 & 500 & 0.02 & 0.85 & 0.12 & 0.95 \\
\hline\hline
\end{tabular}
\end{center}

\vspace{0.4cm}
\noindent
eq.(\ref{sigmatt}), {\it the PGB effects for
$m^\prime_t=5$ GeV in Model B can be marginally detected at the LHC (at
least for $M_{\Pi_t}\alt 800$ GeV and $M_{\Pi^{0a}}\alt 450$ GeV)}.
For $m^\prime_t=15$ GeV, the relative corrections
$\Delta\sigma^{(B2)}/\sigma_0$ are in the range of $(15\--27)\%$
which are larger than the systematic error and the statistical uncertainty.
Thus {\it the PGB effects for $m^\prime_t=15$ GeV in Model B can be
clearly detected at the LHC}. Comparing the cross sections in Table I and 
Table III, we see that the relative differences between Model A and Model B 
at the LHC are $R^{(1)}_{AB}\equiv (\sigma^{(A)}_{t\bar t}
-\sigma^{(B1)}_{t\bar t})/\sigma^{(A)}_{t\bar t}\approx (46\--61)\%$ and
$R^{(2)}_{AB}\equiv (\sigma^{(A)}_{t\bar t}-\sigma^{(B2)}_{t\bar t})/
\sigma^{(A)}_{t\bar t}\approx (40\--53)\%$. These are all much larger
than the systematic error and the statistical uncertainty, so that
{\it Model A and Model B can be clearly distinguished at the LHC}.

As an illustration, the $t\bar t$ invariant mass distributions in Model B for
$M_{\Pi^{0a}}=400$ GeV and $M_{\Pi_t}=500$ GeV at the Tevatron and
the LHC are shown in Fig. 3. Since the width of $\Pi^{0a}$ in Model B depends 
on $m^\prime_t/F_\Pi$ rather than on $m_t/F_\Pi$, {\it the resonance of 
$~\Pi^{0a}$ in Model B is much sharper than that in Model A}. This is a clear 
distinction between Model B and Model A. The width of $\Pi_t$ is very wide due 
to the largeness of $(m_t-m^\prime_t)/F_{\Pi_t}$ (the smallness of 
$F_{\Pi_t}$). Because of the large width of $\Pi_t$, no resonance peak
of $\Pi_t$ can be seen, and the contribution of $\Pi_t$ is just a
slight enhancement of the $M_{t\bar t}$ distribution in a certain region. In 
Fig. 3(b), the solid curve and the dotted curve denote the $M_{t\bar{t}}$ 
distribution with and without the $\Pi_t$ contribution, respectively. From the
difference of these two curves, we can see the effect of the $\Pi_t$ 
contribution.
We see that both the $\Pi^{0a}$ and the $\Pi_t$ contributions look very
different from those of the new heavy vector resonances (with the 
width about $20\%$ of the mass) shown in Ref.\cite{HP}.

\subsection*{C. Results of Model C}

The obtained cross sections in Model C at the Tevatron and the LHC are
listed in Table IV and Table V, respectively. The cross sections
in Table IV are consistent with the CDF and D0 data. In Model C, the decay
constant $F_\Pi$ is much smaller than that in Model B, so that the
$\Pi^{0a}$ and $\Pi^0$ contributions are enhanced\footnote{In this
paper, we have considered the effect of ideal mixing of $\Pi^0$ in
model C, while this effect is not considered in Ref.\cite{hadtt}.}, and
thus the cross sections in Tables IV and V are larger than those in Tables
II and III. 

\null\vspace{0.4cm}
{\small Table IV. Cross sections in Model C at the $\sqrt{s}=2$ TeV Tevatron. 
$\Delta\sigma^{(Ci)}$ denotes the correction to the SM tree-level cross
section $\sigma_0$, and $\sigma^{(Ci)}_{t\bar{t}}=\sigma_0
+\Delta\sigma^{(Ci)}$ is the total cross section. The superscript $i$
denotes the two cases of $m_t^ {\prime}=5$ GeV ($i=1$) and $m_t^\prime=15$
GeV ($i=2$). All masses are in GeV, and all cross sections are in pb.}
\begin{center}
\doublerulesep 0.5pt
\tabcolsep 5pt
\begin{tabular}{cccccc}
\hline\hline
$M_{\Pi_t^0}$ & $M_{\Pi^{0a}}$  & $\Delta\sigma^{(C1)}$ &
$\sigma^{(C1)}_{t\bar{t}}$ & $\Delta\sigma^{C(2)}$ & $\sigma^{(C2)}
_{t\bar{t}}$ \\
\hline
500 & 400 & 0.50 & 8.52 & 3.46 & 11.48 \\
500 & 450 & 0.33 & 8.35 & 1.91 & 9.93 \\
500 & 500 & 0.21 & 8.23 & 0.99 & 9.00 \\
600 & 400 & 0.48 & 8.49 & 3.43 & 11.45 \\
600 & 450 & 0.30 & 8.32 & 1.88 & 9.90 \\
600 & 500 & 0.18 & 8.20 & 0.96 & 8.98 \\
700 & 400 & 0.46 & 8.47 & 3.42 & 11.44 \\
700 & 450 & 0.28 & 8.30 & 1.87 & 9.89 \\
700 & 500 & 0.16 & 8.18 & 0.95 & 8.96 \\
800 & 400 & 0.45 & 8.47 & 3.42 & 11.43 \\
800 & 450 & 0.28 & 8.29 & 1.86 & 9.88 \\
800 & 500 & 0.16 & 8.17 & 0.94 & 8.96 \\
900 & 400 & 0.44 & 8.46 & 3.41 & 11.43 \\
900 & 450 & 0.27 & 8.29 & 1.86 & 9.38 \\
900 & 500 & 0.15 & 8.17 & 0.94 & 8.95 \\
1000 & 400 & 0.44 & 8.46 & 3.41 & 11.43 \\
1000 & 450 & 0.27 & 8.29 & 1.86 & 9.87 \\
1000 & 500 & 0.15 & 8.16 & 0.93 & 8.95 \\
\hline\hline
\end{tabular}
\end{center}

From Table IV we see that, at the tevatron, the relative correction
$\Delta\sigma^{(C1)}/\sigma_0$ is about $(2\--6)\%~$ which is at most
of the same order as the expected systematic error, and 
$\Delta\sigma^{(C2)}/\sigma_0$ is around $~(12\--43)\%~$ which is larger 
than the systematic error and the statistical uncertainty. So that, at the 
Tevatron, {\it the PGB effects in Model C ~for $m^\prime_t=5$ GeV can hardly 
be detected, while those for $m^\prime_t=15$ GeV can be clearly detected}. 
The relative differences 
$R^{(2)}_{CB}\equiv (\sigma^{(C2)}_{t\bar t}-\sigma^{(B2)}_{t\bar t})/
\sigma^{(C2)}_{t\bar t}\approx (9\--26)\%$, so that, for
$m^\prime_t=15$ GeV, {\it Model C can be distinguished from Model B at the 
Tevatron}. However, the relative difference 
$R^{(2)}_{CA}\equiv (\sigma^{(C2)}_{t\bar t}-\sigma^{(A)}_{t\bar t})/
\sigma^{(C2)}_{t\bar t}$ is at most $5\%$, therefore, {\it even for 
$m^\prime_t=15$ GeV, Model C can hardly be distinguished from Model A at the 
Tevatron}.

\vspace{0.4cm}
{\small Table V. Cross sections in Model C at the $\sqrt{s}=14$ TeV LHC. 
$\Delta\sigma^{(Ci)}$ denotes the correction to the SM tree level cross 
sections $\sigma_0$, and $\sigma^{(Ci)}_{t\bar{t}}=\sigma_0
+\Delta\sigma^{(Ci)}$ is the total cross section. The superscript $i$ 
denotes the two cases of $m_t^ {\prime}=5$ GeV ($i=1$) and $m_t^\prime=15$ GeV
($i=2$). All masses are in GeV, and all cross sections are in nb.}
\begin{center}
\doublerulesep 0.5pt
\tabcolsep 5pt
\begin{tabular}{cccccc}
\hline\hline
$M_{\Pi_t^0}$ & $M_{\Pi^{0a}}$ & $\Delta\sigma^{(C1)}$ &
$\sigma^{(C1)}_{t\bar{t}}$ & $\Delta\sigma^{(C2)}$ & $\sigma^{(C2)}
_{t\bar{t}}$ \\
\hline
500 & 400 & 0.22 & 1.04 & 1.64 & 2.47 \\
500 & 450 & 0.20 & 1.02 & 1.35 & 2.18 \\
500 & 500 & 0.16 & 0.99 & 1.02 & 1.85 \\
600 & 400 & 0.20 & 1.03 & 1.63 & 2.45 \\
600 & 450 & 0.19 & 1.01 & 1.34 & 2.17 \\
600 & 500 & 0.15 & 0.98 & 1.01 & 1.84 \\
700 & 400 & 0.19 & 1.02 & 1.62 & 2.45 \\
700 & 450 & 0.18 & 1.00 & 1.33 & 2.16 \\
700 & 500 & 0.14 & 0.97 & 1.00 & 1.83 \\
800 & 400 & 0.19 & 1.02 & 1.62 & 2.44 \\
800 & 450 & 0.17 & 1.00 & 1.33 & 2.15 \\
800 & 500 & 0.14 & 0.96 & 1.00 & 1.82 \\
900 & 400 & 0.19 & 1.01 & 1.61 & 2.44 \\
900 & 450 & 0.17 & 1.00 & 1.33 & 2.15 \\
900 & 500 & 0.13 & 0.96 & 1.00 & 1.82 \\
1000 & 400 & 0.18 & 1.01 & 1.61 & 2.44 \\
1000 & 450 & 0.17 & 0.99 & 1.32 & 2.15 \\
1000 & 500 & 0.13 & 0.96 & 0.99 & 1.82 \\
\hline\hline
\end{tabular}
\end{center}

\null\vspace{0.4cm}

From Table V we see that, at the LHC, the relative corrections 
$\Delta\sigma^{(C1)}/\sigma_0\approx (16\--26)\%$,
$\Delta\sigma^{(C2)}/\sigma_0\approx (120\--200)\%$. These are all much
larger than the systematic error and the statistical uncertainty. So that 
{\it the PGB effects in Model C, for both $m^\prime_t=5$ Gev and 
$m^\prime_t=15$ GeV, can be clearly detected at the LHC}.
Comparing the cross sections in Table V with those in Table I and Table III, we
see that the relative differences are~~ $R^{(1)}_{AC}\equiv 
(\sigma^(A)_{t\bar t}-\sigma^{(C1)}{t\bar t})/
\sigma^{(A)}_{t\bar t}\approx (40\--54)\%$,~~  $R^{(1)}_{CB}\equiv
(\sigma^{(C1)}_{t\bar t}-\sigma^{(B1)}_{t\bar t})/\sigma^{(C1)}_{t\bar t}
\approx (11\--15)\%$,~~  $R^{(2)}_{CA}\equiv (\sigma^{(C2)}_{t\bar t}
-\sigma^{(A)}_{t\bar t})/\sigma^{(C2)}_{t\bar t}\approx (11\--16)\%$, 
~~ $R^{(2)}_{CB}\equiv (\sigma^{(C2)}_{t\bar t}-\sigma^{(B2)}_{t\bar
t})/\sigma^{(C2)}_{t\bar t}\approx (47\--58)\%$. These are all much
larger than the systematic error and the statistical uncertainty. So
that for both $m^\prime_t=5$ GeV and $m^\prime_t=15$ GeV, {\it Model C
can be clearly distinguished from Model A and Model B at the LHC}.

For comparison with Fig. 2 and Fig. 3, the corresponding $t\bar t$ invariant 
mass distributions at the Tevatron and the LHC in Model C are illustrated in 
Fig. 4. We see that {\it the resonances of $\Pi^{0a}$ are significantly wider 
than those in Model B, and clearly narrower than those in Model A} 
because the width of $\Pi^{0a}$ depends on $m^\prime_t/F_\Pi$, and the
values of $F_\Pi$ are very different in Model B and Model C. This character 
shows the clear distinction of the three kinds of TC models. Here we
see again that the $\Pi_t$ contribution does not show up as a resonance
peak, and its effect can be seen from the difference between the curve
with its contribution (the solid curve) and the curve without its
contribution (the dotted curve). The shapes of the $\Pi^{0a}$ and $\Pi_t$
contributions in Model C all look very different from those of the new
heavy vector reaonaces shown in Ref.\cite{HP}.

\null\vspace{0.4cm}
\begin{center}
{\bf IV. CONCLUSIONS}
\end{center}

In this paper, we have studied the pseudo-Goldstone boson contributions
to the $t\bar t$ production cross sections at the Fermilab Tevatron Run
II and the CERN LHC in various technicolor models, and have examined the
possiblity of testing and distinguishing different technicolor models in the 
experiments. We take the Appilquist-Terning one-family walking
technicolor model (Model A), the original topcolor-assisted technicolor
model (Model B), and the topcolor-assisted multiscale technicolor model
(Model C) as three typical examples of technicolor models with and
without topcolor. At the hadron colliders, the $s$-channel
pseudo-Goldstone boson contrbutions described in Fig. 1 dominate. In the
calculation, the MRS set A$^\prime$ parton distribution functions are used to
obtain the $p(\bar p)\to t\bar t$ cross sections, and the
pseudo-Goldstone boson masses $M_{\Pi^{0a}}$ and $M_{\Pi_t}$ are taken
to vary in certain ranges (as discussed in Sec. II) to see the dependence of 
the cross sections on them. The obtained results are compared with the recent 
CDF and D0 data on the $t\bar t$ production cross sections at the Tevatron 
[cf. eq.(\ref{sigmatt})]. It is shown that all the obtained cross
sections at the Tevatron are consistent with the CDF and D0 data.

The results of the calculated cross sections are listed in Table I to
Table V. Considering the expected systematic error at the Tevatron and
the LHC, and assuming a $10\%$ detecting efficiency, we have the following 
conclusions:

\begin{enumerate}
\item Model A can be clearly detected both at the Tevatron and the LHC.
\item In Model B and Model C, the $\Pi^{0a}$ couplings are proportional
to $m^\prime_t$ ($m^\prime_t\ll m_t$) rather than to $m_t$ as in Model A.
Therefore the $\Pi^{0a}$ contributions in Model B and Model C are
significantly reduced relative to Model A. This causes the fact that,
considering the expected systematic error and the statistic uncertainty,
Model B can hardly be detected at the Tevatron, and model C can be
detected at the Tevatron only for large $m^\prime_t$, say $m^\prime_t=15$ GeV. 
The situation is much better for the LHC. Model B with $m^\prime_t=15$ GeV and 
Model C (with $m^\prime_t=5~{\rm GeV~and}~m^\prime_t=15$ GeV) can all 
be clearly detected, and Model B with
$m^\prime_t=5$ GeV can be marginally detected at the LHC.
\item Due to the smallness of $F_{\Pi_t}$ (the largeness of 
$(m_t-m^\prime_t)/F_{\Pi_t}$), the width of the $\Pi_t$ resonance is
very large which causes the fact that the cross sections are not so sensitive 
to the variation of $M_{\Pi_t}$.
\item For the detectable cases, all the three kinds of models can be
experimentally distinguished by the significant differences of their
cross sections. Furthermore, the $\Pi^{0a}$ resonance peaks in the invariant 
mass $M_{t\bar t}$ distributions for the three kinds of models are also very 
different. 
The width of the $\Pi^{0a}$ resonance in the three models are: 
$\Gamma^{(A)}_{\Pi^{0a}}>\Gamma^{(C)}_{\Pi^{0a}}>\Gamma^{(B)}_{\Pi^{0a}}$.
This can serve as a clear distinction between the three kinds of 
models.
\item Comparing the present results with the heavy vector 
resonances (with the width about $20\%$ of the mass) shown in Ref.\cite{HP}, 
we see that the $\Pi^{0a}$ resonances are much sharper and the $\Pi_t$ 
contributions do not show up as resonances. The behavior of the 
present resonances are very different from those heavy vector 
resonances studied in Ref\cite{HP}.

\end{enumerate}

In summary, the PGB effects in $t\bar t$ productions at the LHC provide 
feasible tests of technicolor models including distinguishing different 
typical models. It is complementary to other tests such as the tests studied in
Refs.\cite{HP,TC2-1,DE,RS,test}.

\vspace{0.4cm}

\begin{center}
{\bf Acknowledgment}
\end{center}

This work is supported by the National Natural Science Foundation of China,
the Fundamental Research Foundation of Tsinghua University, and a
special grant from the Ministry of Education of China.


\onecolumn
\begin{figure}[h]

\vspace*{16cm}
\includegraphics{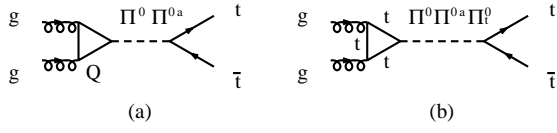}
\vspace{-13cm}
\caption[]{Feynman diagrams for the $s$-channel pseudo-Goldstone boson
contributions to the $p(\bar p)\to t\bar t$ productions: (a) the
techniquark triangle loop contributions, (b) the top quark triangle
loop contributions.}
\end{figure}

\vspace{7cm}
\begin{figure}[h]

\vspace*{10cm}
\includegraphics{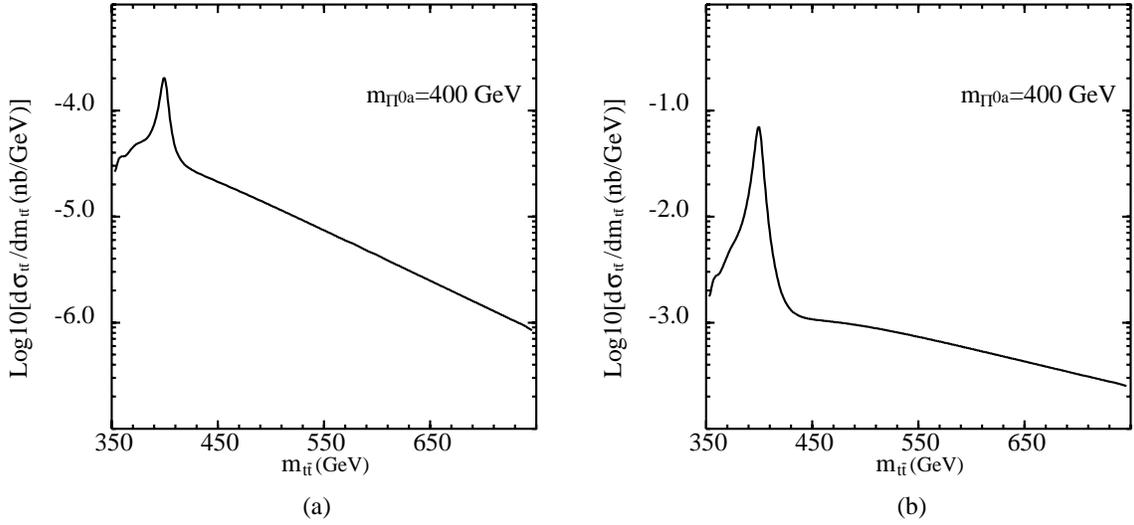}
\vspace{-3cm}
\caption{$t\bar t$ invariant mass distributions for $M_{\Pi^{0a}}=400$
GeV in Model A: (a) at the Tevatron, (b) at the LHC.}
\end{figure}
\newpage
\begin{figure}[h]

\vspace*{10cm}
\includegraphics{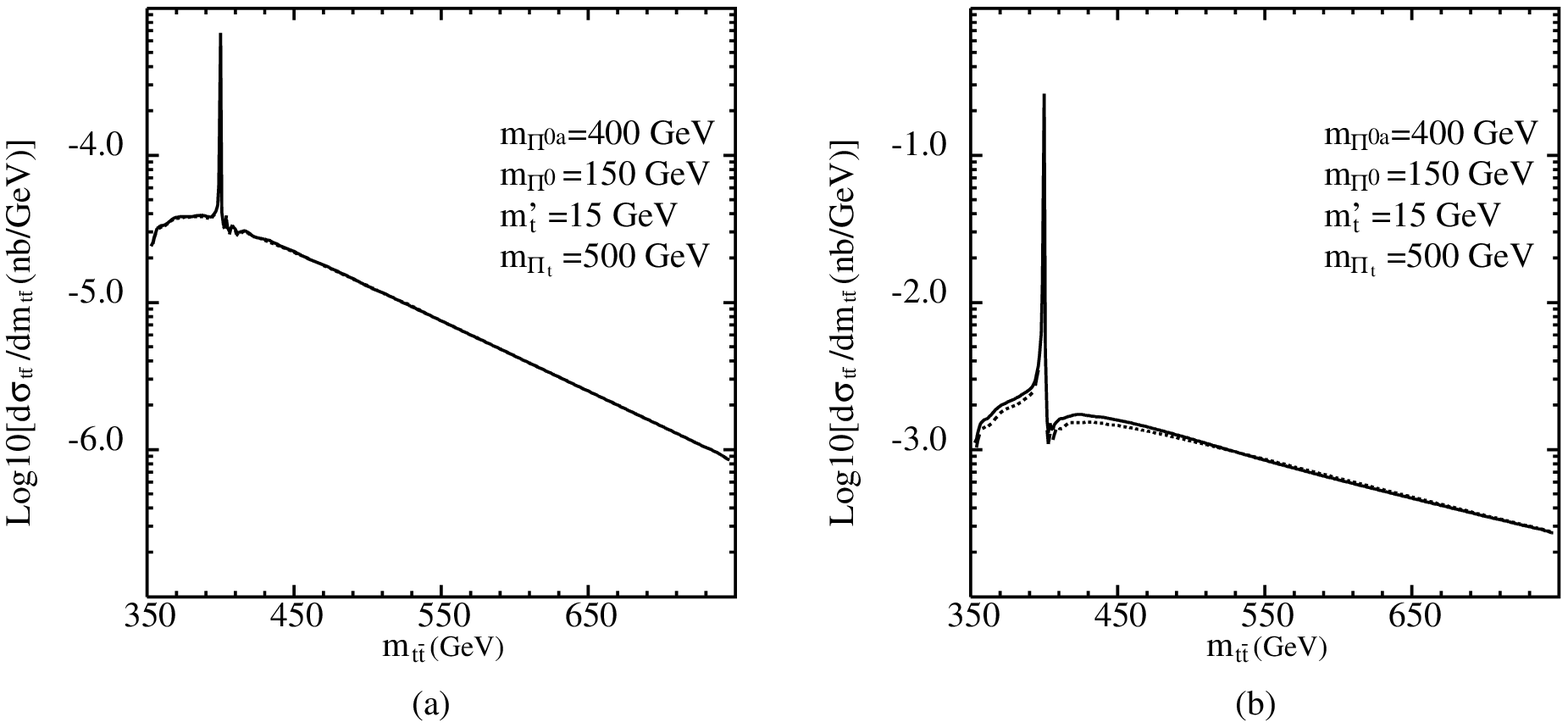}
\vspace{-3cm}
\caption{$t\bar t$ invariant mass distributions for $M_{\Pi^{0a}}=400$ GeV 
and $M_{\Pi_t}=500~{\rm GeV}$ in Model B: (a) at the Tevatron, (b) at the LHC.
The solid and dotted curves in Fig. 3(b) denote the distributions with
and without the $\Pi_t$ contribution, respectively.}
\end{figure}

\vspace{4cm}
\begin{figure}[h]

\vspace*{10cm}
\includegraphics{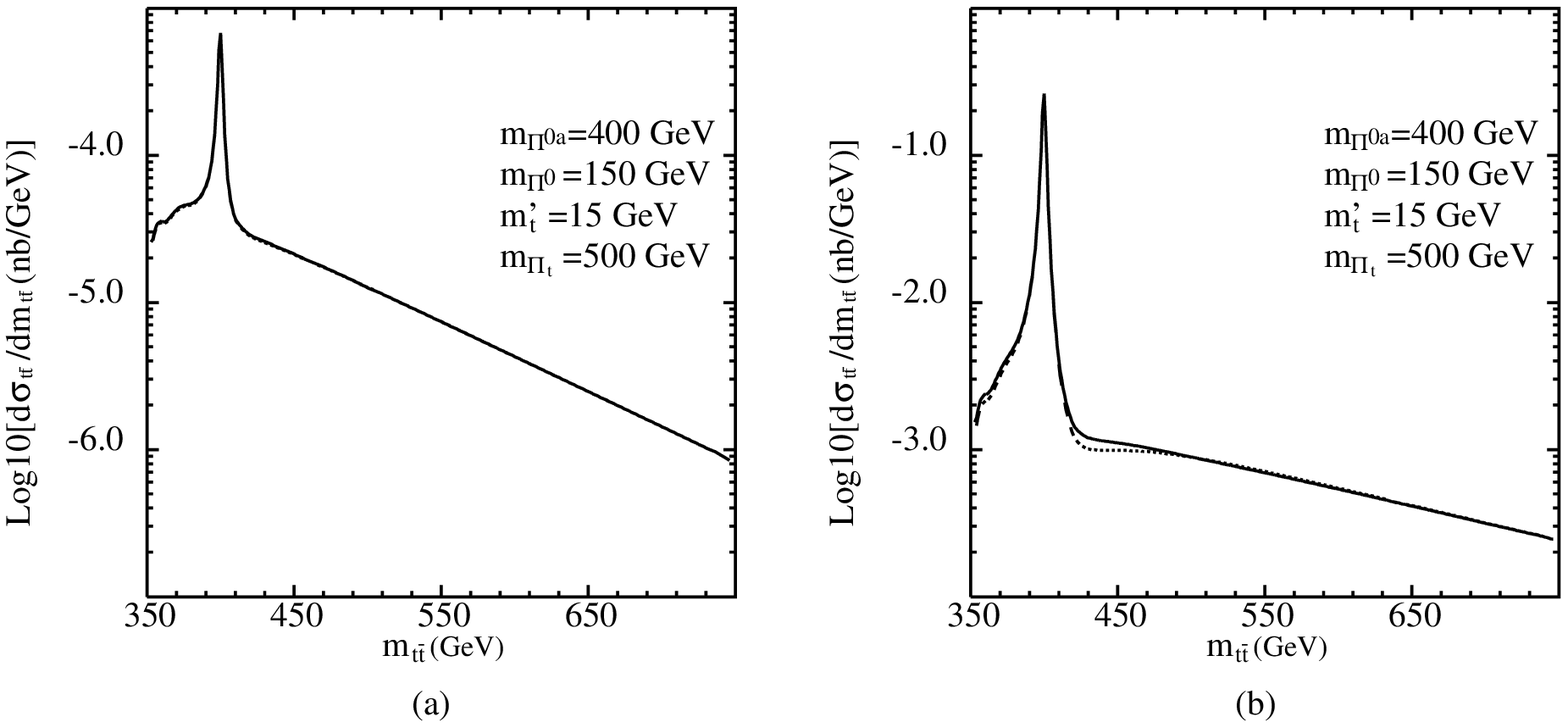}
\vspace{-3cm}
\caption{$t\bar t$ invariant mass distributions for $M_{\Pi^{0a}}=400$ GeV 
and $M_{\Pi_t}=500~{\rm GeV}$ in Model C: (a) at the Tevatron, (b) at the LHC.
The solid and dotted curves in Fig. 4(b) denote the distributions with
and without the $\Pi_t$ contribution, respectively.}
\end{figure}

\end{document}